\documentclass[runningheads]{llncs}
\usepackage[T1]{fontenc}
\usepackage{graphicx}
\usepackage{tikz}
\usetikzlibrary{spy}
\usepackage{amsmath}
\usepackage{amssymb}
\usepackage{bm}

\newcommand{\R}{\mathbb{R}}
\newcommand{\grad}{\bm{\nabla}}
\newcommand{\Div}{\mbox{\rm{\bf{div\,}}}}

\begin{document}
\title{Image Blending with Osmosis}
%
%
\author{Paul Bungert \and
        Pascal Peter \and \\
        Joachim Weickert}

\authorrunning{P. Bungert et al.}
%
\institute{Mathematical Image Analysis Group,
    Faculty of Mathematics and Computer Science,\\ Campus E1.7,
    Saarland University, 66041 Saarbr\"ucken, Germany.\\
    \{bungert, peter, weickert\}@mia.uni-saarland.de}
\maketitle              
\begin{abstract}
Image blending is an integral part of many multi-image applications such as
panorama stitching or remote image acquisition processes. In such scenarios, 
multiple images are connected at predefined boundaries to form a larger image. 
A convincing transition between these boundaries may be challenging, since each 
image might have been acquired under different conditions or even by different 
devices. 

We propose the first blending approach based on osmosis filters. These 
drift-diffusion processes define an image evolution with a non-trivial 
steady state. For our blending purposes, we explore several ways to 
compose drift vector fields based on the derivatives of our input images. These 
vector fields guide the evolution such that the steady state yields a 
convincing blended result. Our method benefits from the well-founded 
theoretical results for osmosis, which include useful invariances under 
multiplicative changes of the colour values. Experiments on real-world 
data show that this yields better quality than traditional gradient 
domain blending, especially under challenging illumination conditions.

\end{abstract}
\section{Introduction}

Image stitching refers to the task of merging multiple images that depict 
different areas of the same scene or object. It has many practical 
applications, in particular for panorama photography \cite{LZPW04} or various
forms of remote image acquisition that create mosaic problems 
\cite{FWFZ19,GMNG09}. In practice,  differing lighting 
conditions in the individual input images may yield large differences in 
brightness and contrast. Therefore, na\"ive stitching as a 
mosaic of the individual input images often creates undesirable boundary 
artefacts. 

Stitching results without visible seams are the objective of so-called 
blending methods. Similar problems exist in image editing, which also requires
seamless merging of content from different sources. In this related field, 
osmosis has yielded excellent results \cite{VHWS13,WHBV13,DMM18}.
This class of filters is inspired by physical processes. In its 
drift--diffusion formulation for visual computing applications, a 
continuous \cite{WHBV13,Sc18} and a discrete \cite{VHWS13} theory have 
been established. 
Moreover, it has also been successfully used for shadow removal 
\cite{WHBV13,DMM18,PCCSW19}, which suggests that it performs well under
large brightness differences. Despite these indications that osmosis filtering 
might also perform well for blending problems, its potential has not been 
investigated so far. Our goal is to address this problem.


\subsection{Our Contributions}

We propose an image blending approach that merges several images seamlessly by 
osmosis. To this end, we use derivative data of the input images to 
define a suitable drift vector field. It guides the osmosis process such that 
it converges to the blending result as a steady state. 

We introduce multiple 
ways to construct these drift vector field and demonstrate that even 
straightforward approaches yield visually highly convincing results. Our 
evaluations on real world and synthetic data reveal that osmosis blending 
benefits from its multiplicative invariance, especially on challenging image 
sets with high contrast differences.


\subsection{Related Work}

Osmosis models in visual computing go back to Weickert et al.~\cite{WHBV13}. 
They have presented a continuous theory along with applications for compact
image representations, shadow removal, and seamless image fusion. The 
continuous theory has been extended by Schmidt~\cite{Sc18}. Vogel et 
al.~\cite{VHWS13} have established a discrete framework and have proposed 
a fast implicit solver. These considerations refer to linear osmosis models,
which already allow a large degree of flexibility.

On the application side, Parisotto et al.~\cite{PCCSW19} have advocated a 
nonlinear variant of osmosis for shadow removal. As a new application 
field for osmosis, Parisotto et al.~\cite{PCD18} proposed the fusion of 
spectral images with linear osmosis, and they also considered nonlinear 
fusion approaches \cite{PCBP+20}.

Osmosis models have several interesting predecessors and share
some conceptual similarities with other methods. A lattice Boltzmann 
model for halftoning by Hagenburg et al.~\cite{HBVW09} can be seen as 
an early nonlinear drift--diffusion process in visual computing.
Outside the field of visual computing, Hagenburg et al.~\cite{HBWV12} 
advocated osmosis models for improving numerical methods for
hyperbolic conservation laws. They used a Markov chain formulation
rather than a drift--diffusion model. In statistical physics, 
drift--diffusion has close ties to the Fokker-Planck equation \cite{Ri84} 
and by that connections to Langevin formulations and the 
Beltrami flow \cite{So01b}.

With its ability to ``integrate'' nonintegrable derivative information 
in a seamless way, osmosis resembles gradient domain methods. They 
have been introduced for shape from shading by Frankot and Chelappa 
\cite{FC88} and have found numerous applications in computer graphics. 
These include tone mapping \cite{FLW02} and image editing~\cite{PGB03}, but
also image blending~\cite{LZPW04}, as we will discuss below.

In contrast to osmosis models that are invariant under multiplicative
rescalings of the pixel values, gradient domain methods are
invariant under additive rescalings. Georgiev's covariant derivative 
approach for image editing~\cite{Ge06} is invariant under multiplicative 
changes, but has not been applied to image blending so far.
Illner and Neunzert~\cite{IN93} have proposed directed diffusion which 
shares conceptual similarities with osmosis in that it converges 
to a so-called background image. However, they did not use it 
for any visual computing application.

Regarding image blending, traditional methods either attempted to find 
optimal boundaries between the input images in order to minimise 
artefacts~\cite{EF01} or directly average information between overlapping 
regions, e.g. with so-called feathering \cite{UES01} or pyramid blending 
\cite{BA83}. The gradient domain method of Levin et al.~\cite{LZPW04} not only 
yielded a significant gain in quality over the image domain method, but also 
has the closest relation to our own osmosis blending. This connection is 
discussed in detail in Section~\ref{sec:integration}. 

While a full review of blending is beyond the scope of this paper, there are 
also approaches with watershed segmentation and graph cuts \cite{GMNG09} or 
wavelets \cite{SHA11}. Others focus on compensating colour differences 
\cite{FWFZ19}, and there are also blending methods based on deep learning 
\cite{WZZH19}. Due to its direct focus on the blending problem itself and 
conceptual relations, we will focus our comparative evaluation on the gradient 
domain method of Levin et al.~\cite{LZPW04}. 


\subsection{Organisation of the Paper}
In Section~\ref{sec:osmosis} we review the theoretical background of osmosis 
process which constitutes the foundation of our blending. Combining this basic 
model with multiple different ways to guide the osmosis-driven propagation 
leads to the blending approaches described in Section~\ref{sec:blending}. We 
compare these methods against each other and gradient domain blending in 
Section~\ref{sec:experiments} and conclude with a discussion and outlook on 
future work in Section~\ref{sec:conclusion}.


\section{Osmosis}
\label{sec:osmosis}


\subsection{Continuous Model}

In our blending setting, we consider nonnegative colour images $\bm f : \Omega 
\rightarrow 
\mathbb{R}^{n_c}_+ $. They map the image domain $\Omega \subset \mathbb{R}^2$ 
to a positive colour domain $\mathbb{R}^{n_c}_+ $ with $n_c$ colour channels 
($n_c=3$ for RGB images, $n_c=1$ for greyscale). We denote individual colour 
channels as $f_i$, i.e. $f_i : \Omega \rightarrow \mathbb{R}_+$.

In addition to the initial image $\bm f$, the osmosis process also relies on a 
given multi-channel \emph{drift vector field} $\bm d: \Omega \rightarrow 
\mathbb{R}^{2 n_c}$. 
For 
each channel $i$ of the image $\bm u(\bm x, 
t)$, the osmosis evolution over time $t$ is described by the initial 
boundary value 
problem 
\begin{align}
    \partial_t u_i = \Delta u_i - \Div(\bm d_i u_i) & &&  \text{on } 
    \Omega 
    \times (0,\infty)\text{,}  \label{eq:ospde}\\ 
    u_i(\bm x,0) = f_i(\bm x) & &&  \text{on } \Omega\text{,} 
    \label{eq:osinitial} \\
   \bm n^\top (\grad u_i - \bm d_i u_i ) = 0 &&&  \text{on } 
   \partial 
    \Omega \times (0,\infty)\text{.}\label{eq:osbound}
\end{align}
The linear partial differential equation (PDE) \eqref{eq:ospde} describes the 
propagation of grey levels over time $t$.

The evolution equation~\eqref{eq:ospde} consists of a diffusion part defined by 
the Laplace 
operator $\Delta u_i = \partial_{xx} 
u_i + \partial_{yy} u_i$ and the drift component  $-\Div(\bm d_i u_i) = 
-\partial_x d_{i,1} u_i - \partial_y d_{i,2} u_i$. 
The drift vector field $\bm d_i \in \R^{2}$ contains two-dimensional drift 
vectors for the colour channel $i$, and $d_{i,1}$ and $d_{i,2}$  denote its 
first 
and second components. It distinguishes osmosis from a pure diffusion process, 
which would lead to a flat steady state for $t \rightarrow \infty$. Instead, 
for osmosis, non-flat steady states are possible. 

In addition to the image $\bm f$ as an initial condition at time $0$ in 
Eq.~\eqref{eq:osinitial}, we define homogeneous Neumann boundary conditions in 
Eq.~\eqref{eq:osbound}. They prevent transport across the image boundaries 
$\partial \Omega$ with outer normal vector $\bm n$.   


\subsection{Theoretical Properties}

Weickert et al.~\cite{WHBV13} have shown several characteristic properties of 
osmosis processes. Let the average colour value $\mu_{f_i}$ of a channel 
$f_i$ be described by
\begin{equation}
\mu_{f_i} = \frac{1}{|\Omega|} \int_\Omega f_i(\bm x) \, 
    d \bm x \, .
\end{equation}
As for diffusion processes, this average is preserved by 
osmosis for all intermediate results $\bm u(\cdot, t)$, 
yielding $\mu_{f_i} = \mu_{u_i(\cdot, t)}$. Furthermore, all intermediate 
results 
retain their nonnegativity, i.e. $u_i(\bm x, t) \geq 0$ for all $\bm x \in 
\Omega$, $i \in \{1,...,n_c\}$, and $t > 0$.

For visual computing purposes, statements on the steady-state for $t 
\rightarrow \infty$ are of particular interest. Let us first consider the 
so-called \emph{compatible} case for osmosis. Here, the drift vector field $\bm 
d$ fulfills
\begin{equation}
    \bm d_i = \frac{\grad v_i}{v_i} \label{eq:canonical}
\end{equation}
for a so-called \emph{guidance image} $\bm v : \Omega \rightarrow 
\mathbb{R}^{n_c}_+$. The corresponding drift vector field $\bm d$ is referred 
to as the \emph{canonical drift vector field} of $\bm v$. 
The steady state $\bm w(\bm x)$ of the osmosis process is given by 
\begin{equation}
    w_i(\bm x) = \frac{\mu_{f_i}}{\mu_{v_i}}  v_i(\bm x)
\end{equation}
for all channels $i$. Thus, the osmosis process converges to the guidance image 
rescaled to the average grey value of the initial image.
    
Note here that the definition of the canonical drift vector field in 
Eq.~\ref{eq:canonical} implies a multiplicative invariance w.r.t. the guidance 
image. This is a very useful property for visual computing applications that 
deal with brightness differences. Therefore, osmosis seems well-suited for 
blending.
 
In our blending scenario, we want to obtain the blended image as the steady 
state of an osmosis process. Since this result is unknown, we also do not have 
access to its canonical drift vector field. Therefore, we need to consider the 
so-called \emph{incompatible case} of osmosis.


\subsection{Incompatible Drift Vector Fields and Gradient Domain Methods}
\label{sec:integration}

Weickert et al.~\cite{WHBV13} also consider the steady state of the osmosis 
evolution on the image $\bm u(\bm x, t)$, which implies $\partial_t u_i = 0$ 
for all channels $i$. Plugging 
this into Eq.~\eqref{eq:ospde}, the steady state result 
$\bm w$ fulfills 
\begin{equation}
    \Delta w_i = \Div(\bm d_i w_i) \, . \label{eq:steady}
\end{equation}
This closely resembles the Poisson equation 
\begin{equation}
    \Delta w_i = \Div{\bm g_i} \label{eq:poisson} \, .
\end{equation}
It arises for so-called gradient domain methods \cite{FC88,PGB03} as a 
necessary condition for finding a minimiser of the energy 
\begin{equation}
    E(u_i) = \int_\Omega |\grad u_i - \bm g_i|^2 \, d \bm x \, . 
    \label{eq:energy}
\end{equation}
For a gradient field $\bm g_i = \grad v_i$ this yields an exact integration, 
and thus $\bm v$ as a result, up to an additive constant. This corresponds to 
the compatible osmosis case. For our blending purposes, it will however be 
vital that osmosis exhibits multiplicative invariances instead.

Minimising the energy from Eq.~\eqref{eq:energy} for a non-integrable vector 
field $\bm g_i$ yields an approximate integration result $\bm u_i$. Similarly, 
for osmosis, we can also calculate a steady state that fulfills 
Eq.~\eqref{eq:steady} for a drift vector field $\bm d_i$ which does not 
correspond to a guidance image. This \emph{incompatible} case allows us to 
design drift vector fields that can be used for blending in 
Section~\ref{sec:blending}.


\begin{figure}[ht]
    \centering
    \includegraphics[width=0.9\textwidth]{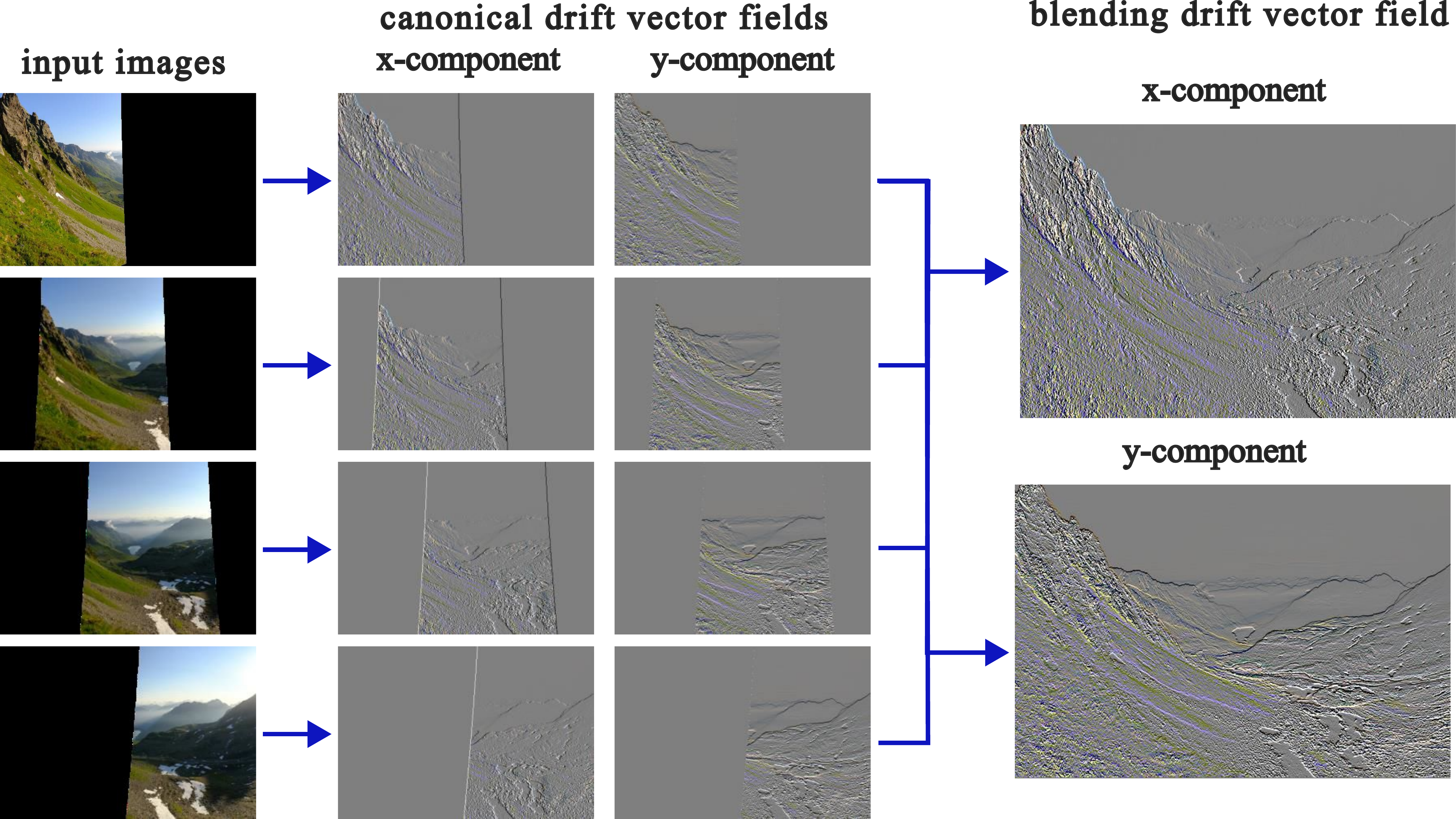}
    \caption{\textbf{Drift Vector Blending.} The canonical drift vector fields 
        for all input images are stitched together at the predetermined 
        seams. The steady state of an osmosis process based on this joint 
        blending 
        drift vector field yields the blended image.}
    \label{fig:osmosis-blending}
\end{figure}


\subsection{Discrete Osmosis}
\label{sec:discrete}

In order to apply osmosis to digital images, we use the discrete 
implementation proposed by Vogel et al.~\cite{VHWS13}. They have shown that the 
theoretical results carry over into the discrete setting and can be implemented 
with appropriate finite difference discretisations.

We are only interested in the steady state, i.e. the solution of the elliptic 
equation \eqref{eq:steady}. Therefore, we use an implicit scheme with a 
stabilised BiCGSTAB solver as described by Meister~\cite{Me15}, since this 
yields faster results compared to an explicit 
formulation \cite{VHWS13}.

Note that in the scheme from~\cite{VHWS13}, the drift vector field is 
discretised on a grid shifted by $h/2$ for a grid size $h$. Therefore, the 
discrete samples coincide with pixel boundaries. This is of particular 
importance for our seam removal approach in Section~\ref{sec:seamremoval}.


\section{Blending with Osmosis}
\label{sec:blending}

In the following we define multiple blending approaches based on the osmosis 
model from Eq.~\eqref{eq:ospde}--\eqref{eq:osbound}. Our blending differs only 
in terms of the drift vector field $\bm d$. For all approaches, we assume that 
the panoramic or mosaic images are already aligned manually or by a suitable 
algorithm.

All following approaches share the same basic problem formulation. Given 
are $n$ aligned, partially overlapping images $\bm v_1, ..., \bm v_n \in 
\R_{+}^{n_x \times n_y \times n_c}$ with spatial resolution $n_x \times n_y$ 
and $n_c$ colour channels. Our approach merges these images into a result $\bm 
w \in \R_{+}^{n_x \times n_y \times n_c}$, the steady state of an 
osmosis process.  As illustrated by Fig.~\ref{fig:osmosis-blending}, the 
aligned input images are padded to the target resolution $n_x \times n_y$. 

Note that according to the properties of osmosis discussed in 
Section~\ref{sec:osmosis}, the steady state is almost completely independent of 
the initial image. Only the average colour value in each channel carries over, 
otherwise the steady state is fully determined by the drift vector 
field. The osmosis process can thus be initialised with a na\"ively 
stitched image or a flat image with the same average colour value. Both 
variants will lead to the same results. Moreover, osmosis preserves 
positivity, but not necessarily the maximum colour values. Therefore, 
we clip the osmosis steady state to the original image range $[0,255]$.


\subsection{Drift Vector Blending}
\label{sec:hard}

The most straightforward approach to osmosis blending is to build a composite 
drift vector field by stitching at a hard seam. To this end, we first 
 compute the canonical drift vector fields $\bm d_1, ...., \bm d_n$ that 
 correspond 
 to the input images $\bm v_1, ..., \bm v_n$ according to 
 Eq.~\eqref{eq:canonical}. 
 
 We split overlapping image parts in the 
 middle, thus partitioning the image into $n$ parts. The corresponding drift 
 vector fields $\bm d_i$ form a partition of the joint drift vector field $\bm 
 d$ as depicted in Fig.~\ref{fig:osmosis-blending}. At partition 
 boundaries, the drift vectors at the pixel boundary overlap (see 
 Section~\ref{sec:discrete}). At such locations, we average the values of both 
 adjacent drift vector fields.
 
Instead of splitting images in the middle, one can also attempt to minimise 
brightness differences between neighbouring pixels with an optimal seam 
algorithm. For our experiments we follow Levin et al.~\cite{LZPW04} and use the 
minimum error boundary cut \cite{EF01}. It computes a seam
between two image areas by minimising the overlapping error in terms of the 
Euclidean distance. The path along the pixel boundaries with the lowest overall 
error is computed with dynamic programming.


\subsection{Seam Removal}
\label{sec:seamremoval}

The blending problem can be also interpreted as a generalisation of shadow 
removal. In both cases, there are multiple regions with different illumination 
and the goal is to fuse them seamlessly into a single image. 

Therefore, we also investigate if established methods for osmosis-based shadow 
removal yield better results than the drift vector stitching. Weickert et 
al.~\cite{WHBV13} compute the canonical drift vector field of the image to be 
edited and modify it by setting the drift vector field to zero at the shadow 
boundaries.

After acquiring seams as in the previous section, we first stitch the images 
together directly in the image domain without performing any blending. This 
allows us to compute the canonical drift vector field of this preliminary 
composite image. The seams coincide with pixel boundaries, which also 
correspond to the locations of the drift vector field in our discretisation 
(see Section~\ref{sec:discrete}). Therefore, we can simply edit the drift 
vector field by setting it to zero at seam locations. 


\begin{figure}[t]
\tabcolsep3pt
\begin{center}
\begin{tabular}{cc|cc}
\multicolumn{2}{c|}{additive changes} &
\multicolumn{2}{c}{multiplicative changes} \\
\begin{tikzpicture}
    [spy using outlines={rectangle, magnification=5, height=1.7cm, 
        width=2.5cm,  connect spies}]
    \node [anchor=south west, inner sep=0pt] {
        \includegraphics[width=0.225\linewidth]{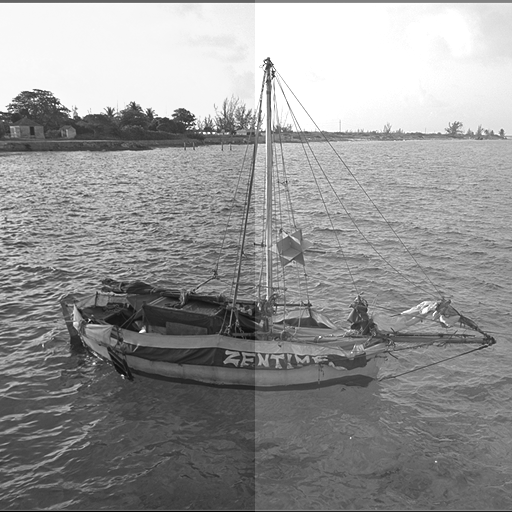}
    } ;         
    \spy [red] on (1.5, 1.7) in node at (1.5,-1);
\end{tikzpicture} &
 \begin{tikzpicture}
    [spy using outlines={rectangle, magnification=5, height=1.7cm, 
        width=2.5cm, 
        connect   spies}]
    \node [anchor=south west, inner sep=0pt] {
        \includegraphics[width=0.225\linewidth]{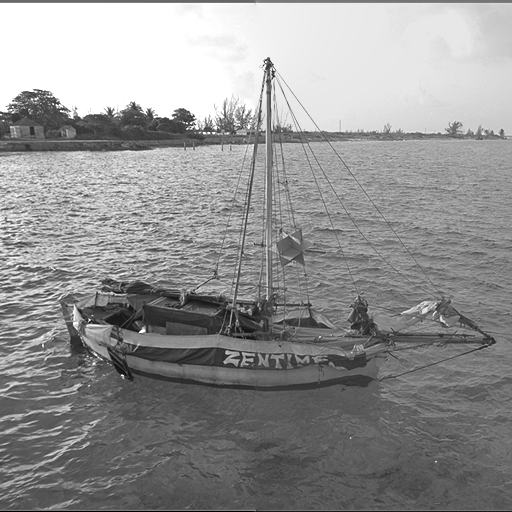}
    } ;
    \spy [red] on (1.5, 1.7) in node at (1.5,-1);
\end{tikzpicture}\hspace*{2mm}&
\hspace{2mm}\begin{tikzpicture}
    [spy using outlines={rectangle, magnification=5, height=1.7cm, 
        width=2.5cm, 
        connect   spies}]
    \node [anchor=south west, inner sep=0pt] {
        \includegraphics[width=0.225\linewidth]{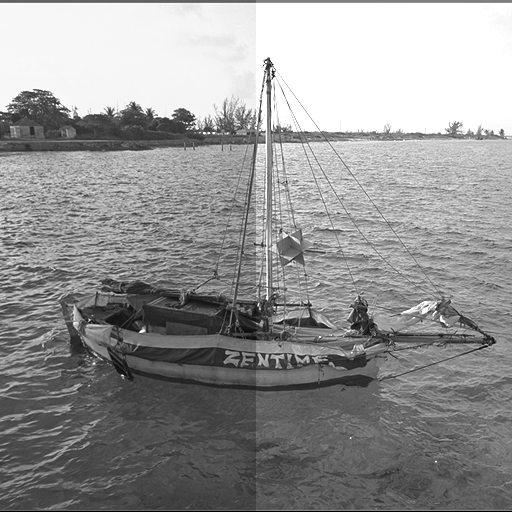}
    } ;         
    \spy [red] on (1.5, 1.7) in node at (1.5,-1);
\end{tikzpicture}&
\begin{tikzpicture}
    [spy using outlines={rectangle, magnification=5, height=1.7cm, 
        width=2.5cm, 
        connect   spies}]
    \node [anchor=south west, inner sep=0pt] {
        \includegraphics[width=0.225\linewidth]{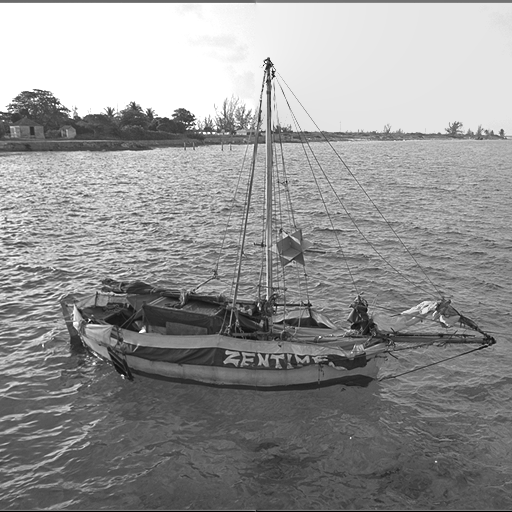}
    } ;
    \spy [red] on (1.5, 1.7) in node at (1.5,-1);
\end{tikzpicture}\\
(a) na\"ive &
(b) osmosis &
(c) na\"ive &
(d) osmosis \\
\end{tabular}
\end{center}%
\caption{\textbf{Invariance Experiments for Osmosis.} Experiments on 
    synthetic examples show that osmosis blending removes seams 
    convincingly 
    for both multiplicative and additive changes.}
\label{fig:invariances}
\end{figure}


\subsection{Alpha Blending with Osmosis}

Levin et al.~\cite{LZPW04} also considered soft seam methods \cite{UES01} for 
gradient 
domain blending. Therefore, we also investigate if this concept can be useful 
in the case of osmosis blending. 

Soft seam approaches do not simply stitch together input data at the seams as 
we did in Section~\ref{sec:hard}. Instead, they perform weighted averaging in 
overlapping regions. Let $\bm d_\ell$ denote the left and $\bm d_r$ the right 
canonical drift vector field in a vertically split overlapping region. 

We blend 
the drift vector fields by location adaptive weighting, i.e. at location $i,j$, 
we get 
\[\bm d_{i,j} = \alpha(i,j) \, \bm d_{\ell,i,j} + (1-\alpha(i,j)) \, \bm 
d_{r,i,j} \, .
\] 
The weight $\alpha$ changes according to its distance to the seam. For vertical 
seams, the shortest distance can be simplified to a description via 
the horizontal index $i$. For our experiments, we define the weight $\alpha$
according to 
\begin{equation}
    \alpha(i,j) = \begin{cases}
        1 &\text{for } i < s_j - w, \\
        \frac{s_j+w-i}{2w} &\text{for } s_j - w \leq i \leq s_j + w, \\
        0 &\text{for } i > s_j + w.
        \end{cases}
\end{equation}
Here, $s_j$ denotes the horizontal position of the seam at the vertical 
position $j$. We weight both fields equally at the 
seam and linearly blend the two drift vector fields inside of the window 
$[s_j-w, s_j+w]$ with size $2w$.   


\begin{figure}[t]
\tabcolsep3pt
\begin{center}
\begin{tabular}{ccc}
\begin{tikzpicture}
    [spy using outlines={rectangle, magnification=5, size = 1.7cm, 
        connect   spies}]
    \node [anchor=south west, inner sep=0pt] {
        \includegraphics[width=0.32\linewidth]{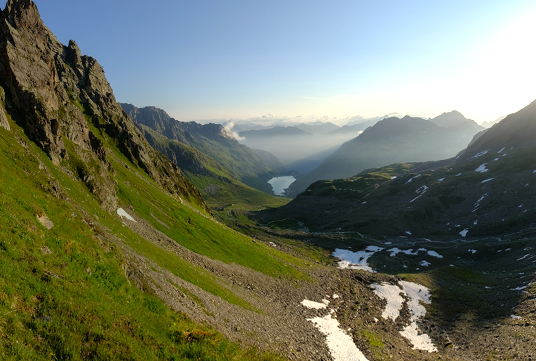}
    } ;
    \spy [red] on (1.3, 1.7) in node at (0.9,-1);
    \spy [red] on  (1.25, 0.2) in node at (3,-1);
\end{tikzpicture}&
\begin{tikzpicture}
    [spy using outlines={rectangle, magnification=5, size = 1.7cm, 
        connect   spies}]
    \node [anchor=south west, inner sep=0pt] {
        \includegraphics[width=0.32\linewidth]{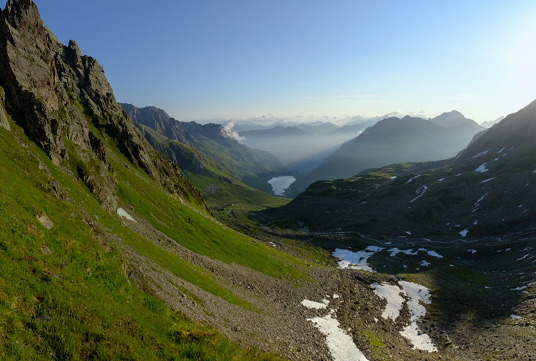}
    } ;
    \spy [red] on (1.3, 1.7) in node at (0.9,-1);
    \spy [red] on  (1.25, 0.2) in node at (3,-1);
\end{tikzpicture}&
 \begin{tikzpicture}
    [spy using outlines={rectangle, magnification=5, size = 1.7cm, 
        connect   spies}]
    \node [anchor=south west, inner sep=0pt] {
        \includegraphics[width=0.32\linewidth]{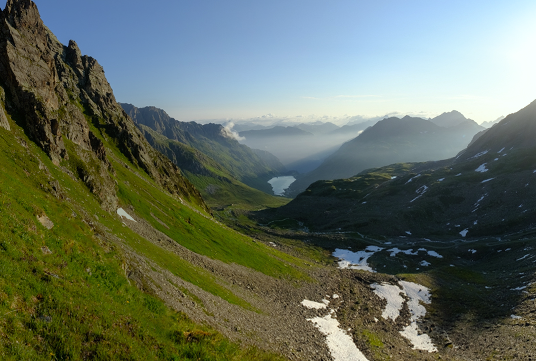}
    } ;
    \spy [red] on (1.3, 1.7) in node at (0.9,-1);
    \spy [red] on  (1.25, 0.2) in node at (3,-1);
\end{tikzpicture}\\
(a) drift vector blending &
(b) seam removal &
(c) alpha blending \\
\end{tabular}
\end{center}  
\caption{\textbf{Comparison of Osmosis Blending Approaches.} The simple 
    direct blending yields results that are just as convincing as seam removal. 
    Alpha blending leads to blurry seams close to the 
    camera, where alignment is imperfect.}
\label{fig:osmosis-approaches}
\end{figure}


\begin{figure}[t]
\tabcolsep3pt
\begin{center}
\begin{tabular}{ccc}
\begin{tikzpicture}
    [spy using outlines={rectangle, magnification=5, size = 1.7cm, 
        connect   spies}]
    \node [anchor=south west, inner sep=0pt] {
        \includegraphics[width=0.32\linewidth]{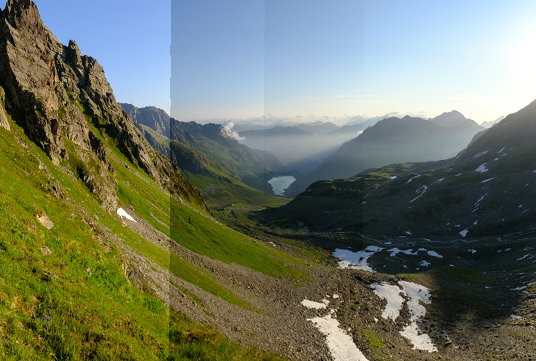}
    } ;
    \spy [red] on (1.3, 1.7) in node at (0.9,-1);
    \spy [red] on  (1.25, 0.2) in node at (3,-1); 
\end{tikzpicture}&
\begin{tikzpicture}
    [spy using outlines={rectangle, magnification=5, size = 1.7cm, 
        connect   spies}]
    \node [anchor=south west, inner sep=0pt] {
        \includegraphics[width=0.32\linewidth]{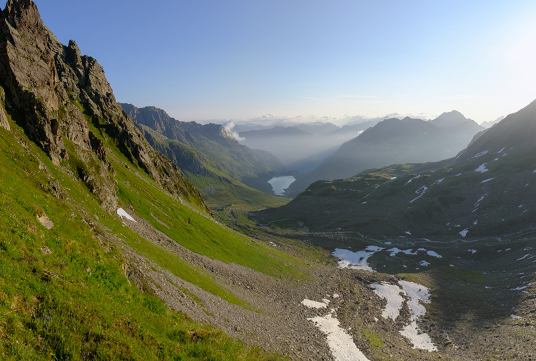}
    } ;    
    \spy [red] on (1.3, 1.7) in node at (0.9,-1);
    \spy [red] on  (1.25, 0.2) in node at (3,-1);
\end{tikzpicture}&
\begin{tikzpicture}
    [spy using outlines={rectangle, magnification=5, size = 1.7cm, 
        connect   spies}]
    \node [anchor=south west, inner sep=0pt] {
        \includegraphics[width=0.32\linewidth]{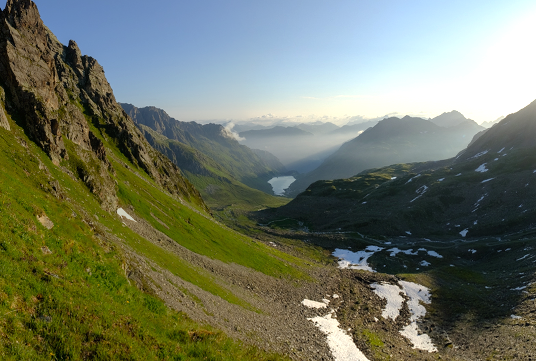}
    } ;
    \spy [red] on (1.3, 1.7) in node at (0.9,-1);
    \spy [red] on  (1.25, 0.2) in node at (3,-1);
\end{tikzpicture}\\
(a) na\"ive blending &
(b) gradient domain &
(c) osmosis \\
\end{tabular}
\end{center}
\caption{\textbf{Plausibility Check: Low Brightness Differences.} As 
    expected, 
    on simple sequences with moderate brightness differences, osmosis and 
    gradient 
    domain blending both perform well. But already here, a slightly better 
    preservance of contrast hints at the robustness of osmosis blending.
    \label{fig:plausibility}}
\end{figure}


\section{Experiments}
\label{sec:experiments}

All real-world images in the evaluation have been aligned and warped with the 
open source tool Hugin  
2021.0.0\footnote{\url{https://hugin.sourceforge.io/}}.
Osmosis was implemented with an implicit scheme and a BiCGSTAB solver as 
proposed by Vogel et al.~\cite{VHWS13}. For each step with a time step 
size of $\tau=10^5$ 
this solver is stopped if the relative Euclidean norm of the residual drops 
below $10^{-9}$. We also determine the number of time steps on a relative 
Euclidean norm on the steady state equation~\eqref{eq:steady}, requiring a 
decay 
of $\| \Delta u_i - \Div(\bm d_i u_i)\|_2$ by more than a factor $10^{-6}$ 
compared to the initialisation.

In addition to evaluating different osmosis blending variants against each 
other, we compare against gradient domain blending. Here, gradient fields of 
the input images are stitched together with the same seams as our osmosis 
method to allow a direct comparison between the properties of these approximate 
integration methods (see Section~\ref{sec:integration}). This corresponds to 
algorithm GIST2 of Levin et al.~\cite{LZPW04}, a widely known approach that is 
conceptually closest to our own.


\subsection{Invariances of Osmosis Blending in Practice}

In Section~\ref{sec:osmosis} we discussed the multiplicative 
invariance for the compatible osmosis case. This is the dominating type of 
illumination changes in blending, and we verify that this carries over 
to the blending application. Moreover, we also investigate the impact of  
additive changes and compare against na\"ive stitching. These are less common, 
but in practice, more complex mixtures of lighting changes than pure 
multiplicative ones can occur.
Therefore we created two synthetic test cases for the image \emph{boat} in 
Fig.~\ref{fig:invariances}. For the additive change in 
Fig.~\ref{fig:invariances}(a), 
each pixel value in the left image half was increased by $30$, and in 
Fig.~\ref{fig:invariances}(c), each pixel was 
multiplied by $1.3$. In both cases, the 
results were clipped at $255$. Osmosis blending is robust under both types of 
brightness changes.


\subsection{Comparing Variants of Osmosis Blending}

On the \emph{mountain} sequence from Fig.~\ref{fig:osmosis-blending} we compare 
our three osmosis blending  approaches from
Section~\ref{sec:blending}. In
Fig.~\ref{fig:osmosis-approaches}, all three approaches equilibrate 
brightness differences well. Drift vector blending and seam removal also 
produce no visible artefacts at the seams. Only alpha blending  
leads to blur at stitching boundaries. Due to  its simplicity and 
quality,  drift 
vector blending is our method of choice.


\subsection{Plausibility Check: Low Brightness Differences}

As a plausibility check, we first investigate the \emph{mountain} sequence with
visible, but fairly low brightness differences for na\"ive blending in 
Fig.~\ref{fig:plausibility}. As expected, gradient domain 
blending yields good results without visible seams. Osmosis offers similar 
quality with a slightly better preservation of high 
contrast, such as the shadow of the mountains in the right half of the image.


\begin{figure}[t]
\tabcolsep3pt
\begin{center}
\begin{tabular}{ccc}
\begin{tikzpicture}
    [spy using outlines={rectangle, magnification=2.5, size = 1.7cm, 
        connect   spies}]
    \node [anchor=south west, inner sep=0pt] {
        \includegraphics[width=0.32\linewidth]{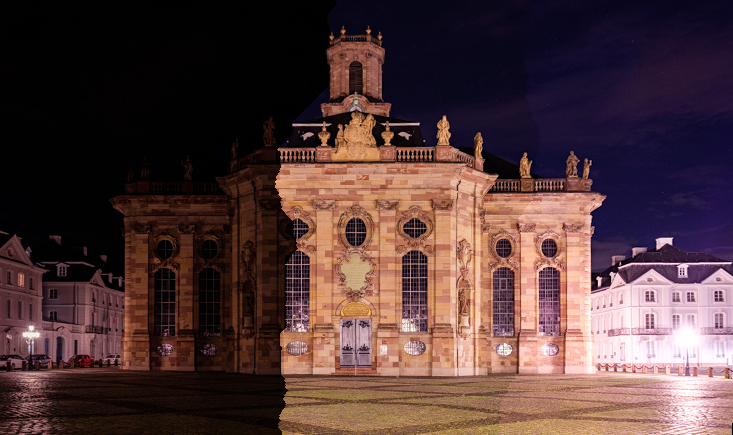}
    } ;
    \spy [red] on (0.5, 0.6) in node at (0.9,-1);
    \spy [red] on  (1.5, 0.4) in node at (3,-1);
\end{tikzpicture}&
\begin{tikzpicture}
    [spy using outlines={rectangle, magnification=2.5, size = 1.7cm, 
        connect   spies}]
    \node [anchor=south west, inner sep=0pt] {
        \includegraphics[width=0.32\linewidth]{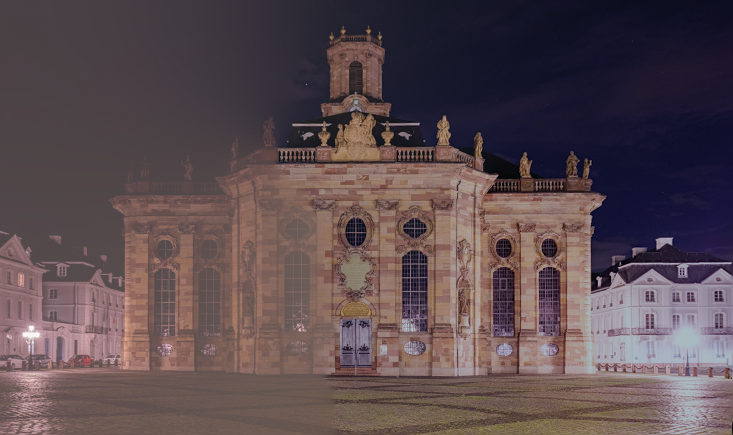}
    } ;
    \spy [red] on (0.5, 0.6) in node at (0.9,-1);
    \spy [red] on  (1.5, 0.4) in node at (3,-1);
\end{tikzpicture}&
\begin{tikzpicture}
    [spy using outlines={rectangle, magnification=2.5, size = 1.7cm, 
        connect spies}]
    \node [anchor=south west, inner sep=0pt] {
        \includegraphics[width=0.32\linewidth]{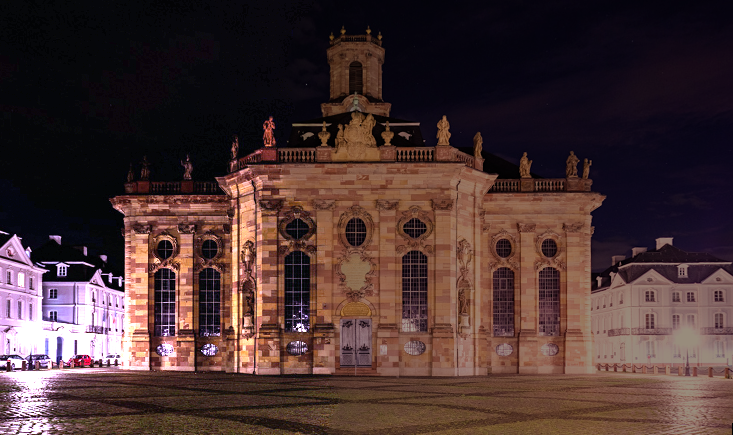}
    } ;
    \spy [red] on (0.5, 0.6) in node at (0.9,-1);
    \spy [red] on  (1.5, 0.4) in node at (3,-1);
\end{tikzpicture}\\
(a) na\"ive blending &
(b) gradient domain &
(c) \textbf{osmosis} \\[3mm]
\begin{tikzpicture}
    [spy using outlines={rectangle, magnification=2.5, size = 1.7cm, 
        connect   spies}]
    \node [anchor=south west, inner sep=0pt] {
        \includegraphics[width=0.32\linewidth]{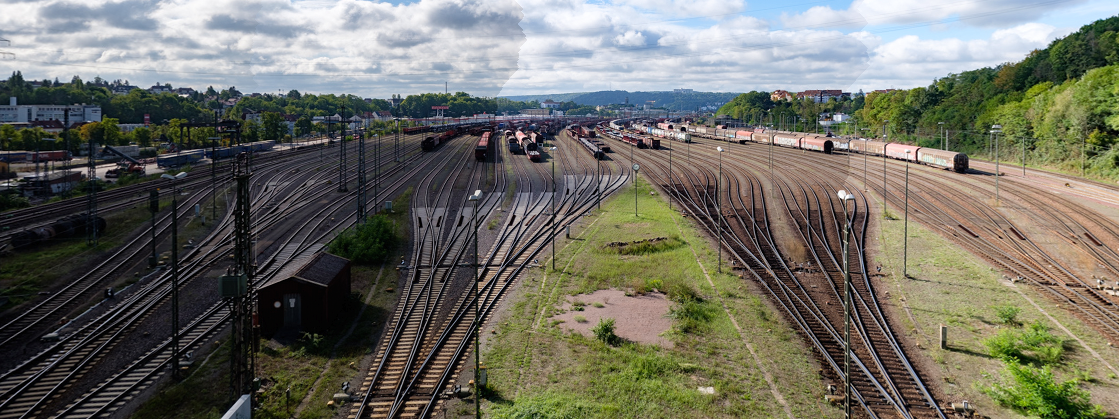}
    } ;
    \spy [red] on (0.5, 0.4) in node at (0.9,-1);
    \spy [red] on (1.7, 1.1) in node at (3,-1);
\end{tikzpicture}&
 \begin{tikzpicture}
    [spy using outlines={rectangle, magnification=2.5, size = 1.7cm, 
        connect   spies}]
    \node [anchor=south west, inner sep=0pt] {
        \includegraphics[width=0.32\linewidth]{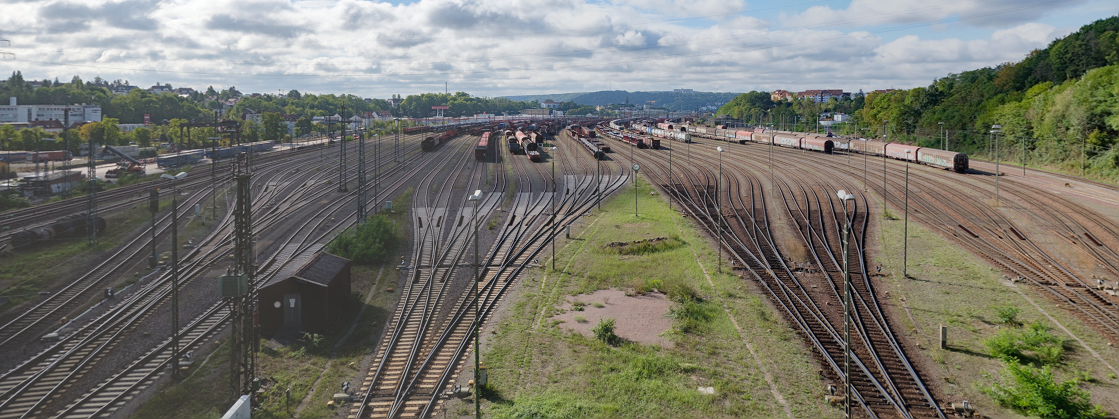}
    } ;
    \spy [red] on (0.5, 0.4) in node at (0.9,-1);
    \spy [red] on (1.7, 1.1) in node at (3,-1);
\end{tikzpicture}&
\begin{tikzpicture}
    [spy using outlines={rectangle, magnification=2.5, size = 1.7cm, 
        connect   spies}]
    \node [anchor=south west, inner sep=0pt] {
        \includegraphics[width=0.32\linewidth]{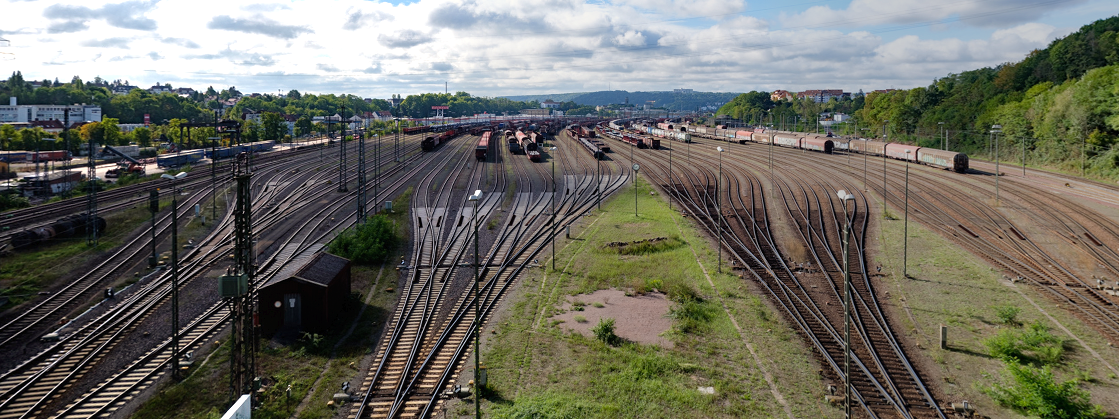}
    } ;
    \spy [red] on (0.5, 0.4) in node at (0.9,-1);
    \spy [red] on (1.7, 1.1) in node at (3,-1);
\end{tikzpicture}\\
(c) na\"ive blending &
(d) gradient domain &
(e) \textbf{osmosis} \\
\end{tabular}
\end{center}
\caption{\textbf{High Brigthness Differences.} We perform optimal seam 
blending of  \emph{building} and  \emph{tracks} 
sequences. On these challenging sequences, osmosis preserves the image 
contrast significantly more convincingly due to its multiplicative 
invariance.}
\label{fig:high-contrast}
\end{figure}


\subsection{High Brightness Differences}

We also evaluate our  blending on two image sets with extreme brightness 
changes. These experiments reveal the differences between our osmosis blending 
and the 
gradient domain approach. For the first set, \emph{building}, illumination 
changes were synthetically  created in Adobe Lightroom v5.5 by darkening one 
half of the input images and 
brightening the other (by 1.5 stops). 
Fig.~\ref{fig:high-contrast}(a)--(c) show that with
optimal seam blending, both the gradient domain approach and osmosis remove the 
seams. However, only osmosis is able to maintain or even enhance the contrast 
in the darkened image parts on the left-hand side. This highlights how the 
multiplicative invariance of osmosis has a significant practical impact. 

The real world sequence \emph{tracks} in 
Fig.~\ref{fig:high-contrast}(d)--(e) contains naturally
occurring large brightness differences. Again, osmosis preserves the contrast 
better than gradient domain blending and yields a considerably more vivid 
result.


\section{Conclusions}
\label{sec:conclusion}

We have proposed the first osmosis model for image blending. Our investigation
has shown that already simple stitching of drift vector fields yields excellent 
results without any visible seams. The natural invariances of the osmosis 
filter take care of multiplicative brightness differences without the need of 
any further processing such as alpha blending.

In particular for challenging image sequences with large 
brightness changes, osmosis clearly produces superior results compared to 
gradient domain methods.

 This application highlights the practical value of the 
theoretical properties provided by osmosis filtering. In the future, we intend 
to leverage these strengths for additional applications in computer vision and 
computer graphics.


%
%
%
\bibliographystyle{splncs04}
\bibliography{bib.bib}
\end{document}